\documentclass[pre]{revtex4}
\usepackage{graphicx}
\begin{document}
\title{Reaction-Diffusion System in a Vesicle with Semi-Permeable Membrane}
\author{Hidetsugu Sakaguchi}
\address{Department of Applied Science for Electronics and Materials,\\
Interdisciplinary Graduate School of Engineering Sciences,\\
Kyushu University, Kasuga, Fukuoka 816-8580, Japan}
\begin{abstract}
 We study the Schl\"ogl model in a vesicle with semi-permeable membrane. 
The diffusion constant takes a smaller value in the membrane region, which prevents the outflow of self-catalytic product. A nonequilibrium  state is stably maintained inside of the vesicle. Nutrients are absorbed and waste materials are exhausted through the membrane by diffusion. It is interpreted as a model of primitive metabolism in a cell.
\end{abstract}
\maketitle
\section{Introduction}
Pattern formation in reaction-diffusion systems is actively studied as a theme of nonequilibrium sciences \cite{rf:1}. Turing type patterns \cite{rf:2}, target patterns and spiral patterns \cite{rf:3} are found in several reaction diffusion systems.  More complicated patterns such as localized patterns \cite{rf:4} and the splitting \cite{rf:5,rf:6} were also found experimentally and theoretically.  The reaction-diffusion systems have become an important model for various nonlinear processes in biological systems. Most reaction-diffusion systems are set up to be as homogeneous as possible to simplify the phenomena and the analyses. Recently, reaction-diffusion systems in more complicated media such as membranes and microemulsions have been studied \cite{rf:7,rf:8}. 

On the other hand, Oparin considered "corecervate" as an origin of the cell, which played an important role of prebiotic chemical evolution \cite{rf:9}. He pointed out the importance of chemical reactions confined in the cellular structure. Dyson pointed out the importance of the jump process to a highly activated state in a certain bistable system as the origin of life \cite{rf:10}. Recently, Noireaux and Libchaber study a cell-like bioreactor using a phospholipid vesicle \cite{rf:11}.  These researchers consider that chemical reactions in a confined cell-like structure  are an important step to life.

A cell is separated from the outside region by a membrane, and complicated chemical reactions occur only inside the cell.  Some materials and reactants are transported through the membrane. 
The consumption of nutrient and doing away of waste matter through the membrane is a basic process of metabolism, and then highly active nonequilibrium state is maintained inside of the cell.  Inside and outside of the cell, materials are in the aqueous solution, but the membrane is composed of lipid. Therefore, the transport coefficients are different at the membrane. 
To understand the confined or localized nonequilibrium states accompanying the transport of materials through the membrane qualitatively, we will study a simple reaction-diffusion system called the Schl\"ogl model in a vesicle with a semi-permeable membrane.  We do not consider complicated reactions including DNA, RNA and so on, 
and  do not consider active transport through the membrane, so it might be too simple as a model of a living cell. 
That is, our model is not a realistic model, but a toy model to consider "metabolism"  from a view point of dissipative structure far from equilibrium.

\section{Localized nonequilibrium state in the Schl\"ogl model}
The Schl\"ogl model is a set of chemical reactions including three kinds of chemicals U,V and W \cite{rf:12} represented as 
\[
{\rm V}+{\rm 2U}\rightleftharpoons {\rm 3U},\; {\rm U}\rightleftharpoons {\rm W}.\]
We interpret the three chemicals V,U and W respectively as nutrient material, self-catalytic product and waste material.
The reaction-diffusion equations for $u=[{\rm U}]$,$v=[{\rm V}]$ and $w=[{\rm W}]$ are expressed as 
\begin{eqnarray}
\frac{\partial u}{\partial t}&=&k_1vu^2-k_2u^3-k_3u+k_4w+D_u\nabla^2 u,\nonumber\\
\frac{\partial v}{\partial t}&=&-k_1vu^2+k_2u^3+D_v\nabla^2 v,\nonumber\\
\frac{\partial w}{\partial t}&=&k_3u-k_4w+D_w\nabla^2w.
\end{eqnarray}
where $k_1,k_2,k_3$  and$k_4$ are rate constants respectively for the reactions V$+$2U$\rightarrow$3U, 3U$\rightarrow$V$+$2U, U$\rightarrow$W and W$\rightarrow$ U, and $D_u,\,D_v, \,D_w$ are diffusion constants for the materials. In usual reaction-diffusion equations, the diffusion constants are assumed to be uniform in space. However, we assume a spherical vesicle surrounded by a membrane as a model of a primitive cell. The diffusion constants inside of the vesicle are different from the bulk region.  We assume non-uniform diffusion constants as 
\begin{eqnarray}
D_u(r)&=&D_{u1}, \;\;\ {\rm for}\;\; r<r_0,\; r>r_0+d,\nonumber\\
 &=&D_{u0} \;\;\;{\rm for}\;\; r_0\le r \le r_0+d,\nonumber\\
D_v(r)&=&D_{v1}, \;\;\ {\rm for}\;\; r<r_0,\; r>r_0+d,\nonumber\\
 &=&D_{v0} \;\;\;{\rm for}\;\; r_0\le r \le r_0+d,\nonumber\\
D_w(r)&=&D_{w1}, \;\;\ {\rm for}\;\; r<r_0,\; r>r_0+d,\nonumber\\
 &=&D_{w0} \;\;\;{\rm for}\;\; r_0\le r \le r_0+d,
\end{eqnarray}
where $r$ is the distance from the center of a spherical cell. 
The value $r_0+d$ denotes the radius of the cell, and $d$ is the thickness of the membrane.  
The diffusion constants are uniform inside and outside of the cell. 
The diffusion constants inside of the cell might be different those outside of the cell, but we assume the same values for the sake of simplicity.  In all simulations in this paper, we have assumed $D_{u1}=D_{v1}=D_{w1}=1$.    
The diffusion constants are assumed to take  smaller values in the membrane region $r_0\le r \le r_0+d$. That is, the transport of materials by the diffusion is assumed to be harder in the membrane region. 

If the rate constants $k_1\sim k_4$ are assumed to be all 1 for simplicity, and the spherical symmetry is further assumed, the model equation is simplified as  
\begin{eqnarray}
\frac{\partial u}{\partial t}&=&vu^2-u^3-u+w+\frac{1}{r^2}\frac{\partial}{\partial r}\left (D_ur^2\frac{\partial u}{\partial r}\right ),\nonumber\\
\frac{\partial v}{\partial t}&=&-vu^2+u^3+\frac{1}{r^2}\frac{\partial}{\partial r}\left (D_vr^2\frac{\partial v}{\partial r}\right ),\nonumber\\
\frac{\partial w}{\partial t}&=&u-w+\frac{1}{r^2}\frac{\partial}{\partial r}\left (D_wr^2\frac{\partial w}{\partial r}\right ).
\end{eqnarray}
The diffusion constants are given by eq.~(2). 
There is a thermal equilibrium state in eq.~(3), because the chemical reactions are reversible. It is an important point different from the usual reaction diffusion equation, in which some irreversible reactions are assumed in most cases. The equilibrium state is characterized by the condition $u=v=w$. 
In our simulation, we have imposed fixed boundary conditions of     
$v=v_0$ and $w=0$ at a radius $r=R=20$ outside of the cell, and no-flux boundary conditions of $\partial u/\partial r=0$ at $r=R=20$, and $\partial u/\partial r=\partial v/\partial r=\partial w/\partial r=0$ at $r=0$.  The radius $r_0$ and the thickness $d$ are fixed to be $r_0=2$ and $d=1$ for most cases.

\begin{figure}[t]
\begin{center}
\includegraphics[height=4.5cm]{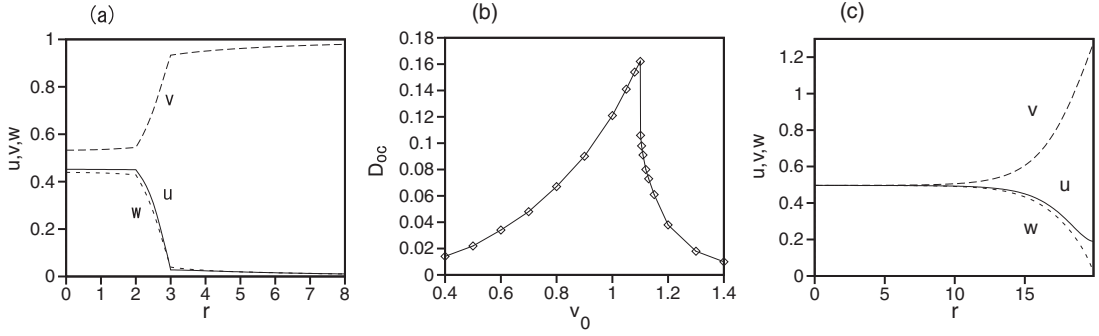}
\end{center}
\caption{(a) Localized solution $u(r),v(r)$ and $w(r)$ to eq.~(3) for $v_0=1,\, r_0=2,\,d=1$ and $D_{u0}=0,\,D_{v0}=D_{w0}=0.05$.  (b) Critical value $D_{0c}$ for the localized state as a function of $v_0$. (c) Extended solution $u(r),v(r)$ and $w(r)$ for $D_0=0.05$ and $v_0=1.3$.}
\label{f1}
\end{figure}
\begin{figure}[t]
\begin{center}
\includegraphics[height=5.5cm]{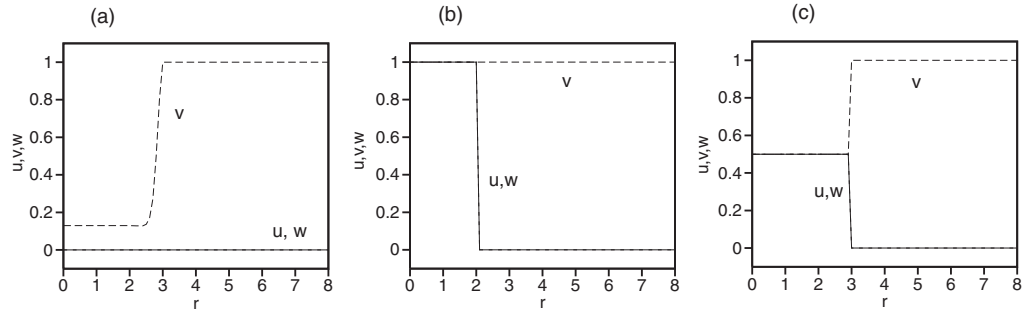}
\end{center}
\caption{(a) Stationary solution $u(r),v(r)$ and $w(r)$ to eq.~(3) for $v_0=1,\, r_0=2,\,d=1$ and $D_{u0}=0.01,\,D_{v0}=0,\,D_{w0}=0.01$ and $D_{u1}=D_{v1}=D_{w1}=1$. (b) Stationary solution $u(r),v(r)$ and $w(r)$ for $D_{u0}=0,\,D_{v0}=0.01,\,D_{w0}=0$ and $D_{u1}=D_{v1}=D_{w1}=1$. (c) Stationary solution $u(r),v(r)$ and $w(r)$ for $D_{u0}=D_{v0}=D_{w0}=0$ and $D_{u1}=D_{v1}=D_{w1}=1$. The initial values are $u(r)=0.6,v(r)=0.5, w(r)=0.4$ for $r<r_0+d$ and $u(r)=v(r)=0, w(r)=v_0=1$ for $r>r_0+d$.}
\label{f1}
\end{figure}
\begin{figure}[t]
\includegraphics[height=4.5cm]{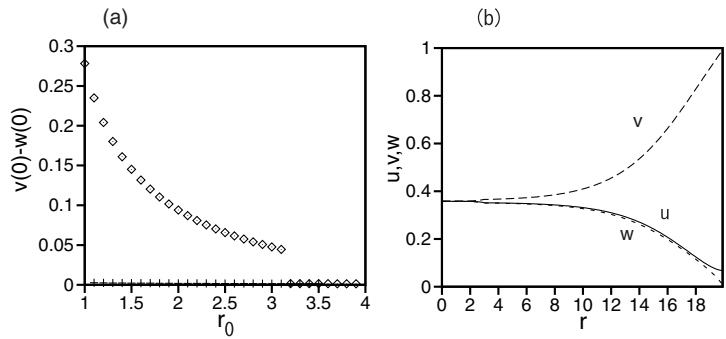}
\caption{(a) Concentration difference $v(0)-w(0)$ as a function of the cell size $r_0$. The upper branch denotes the localized nonequilibrium state and the lower branch denotes an extended nearly-equilibrium solution. (b) Extended nearly-equilibrium solution at $D_{u0}=0,\,D_{v0}=D_{w0}=0.05,\,v_0=1$ and $r_0=2$.}
\label{f2}
\end{figure}\begin{figure}[t]
\begin{center}
\includegraphics[height=4.5cm]{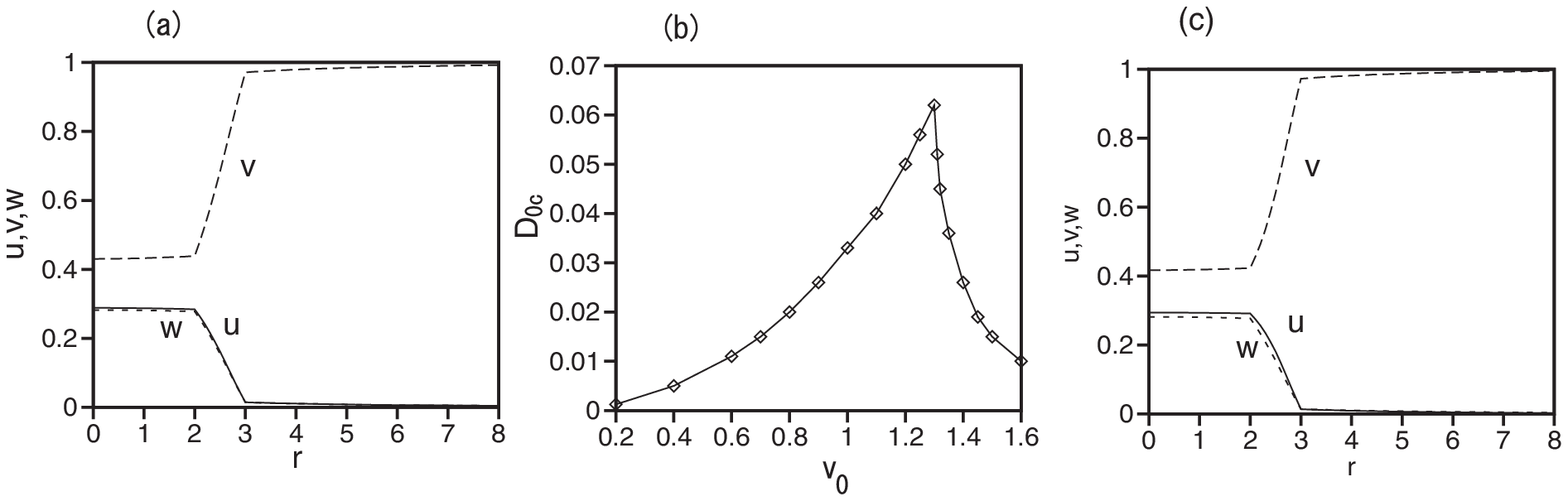}
\end{center}
\caption{(a) Localized solution $u(r),v(r)$ and $w(r)$ to eq.~(3) for $v_0=1,\, r_0=2,\,d=1$ and $D_{u0}=D_{v0}=D_{w0}=0.02$.  (b) Critical value $D_{0c}$ for the localized state as a function of $v_0$. (c) Localized solution of $u(r)$ to eq.~(5) at $\alpha=0.979,\,D_0=0.02$ and $v_0=1$, and $w=\alpha u$ and $v=v_0-u-w$ are also plotted.}
\label{f1}
\end{figure}
The fixed boundary condition $v=v_0\ne 0$ at $r=R$ generates a nonequilibrium state. The completely equilibrium state satisfies $u=v=w$, and it is attained only for the case of $v_0=w_0=0$. If $v_0$ is not zero, the system is deviated from the equilibrium state. There is another uniform solution $u=w=0$ and $v=v_0$. 
The eigenvalues of this uniform state for the uniform perturbation with wavenumber $k=0$ is -2,0 and  0. It implies that the uniform solution is marginally stable. 
Another localized solution can appear. Firstly, we show numerical results in case of $D_{u1}=D_{v1}=D_{w1}=1,\,D_{u0}=0$ and $D_{v0}=D_{w0}=D_0$. 
Here, we consider a situation that the self-catalytic product U cannot diffuse into the membrane and therefore $D_{u0}=0$, assuming that U is a material such as an enzyme of protein and the diffusion hardly occurs through the membrane. 
Numerical simulation was performed with the Euler method with mesh sizes $\Delta t=0.001$ and $\Delta r=0.1$. Figure 1(a) is a stationary state for  $D_{u0}=0, D_{v0}=D_{w0}=0.05$ and $v_0=1$. The nutrient V is absorbed through the membrane $r_0<r<r_0+d$, the concentration of the self-catalytic product U is highly maintained inside of the cell, and the waste material W diffuses out from the membrane and it is absorbed at $r=R$. The equilibrium condition $u=v=w$ is not satisfied inside of the cell. Nonequilibrium localized  state is maintained inside of the cell, owing to the inward flow of the nutrient and the outward flow of the waste material.  This can be interpreted as a simple and primitive model of metabolism. The self-catalytic product U is not transported by diffusion through the membrane because of $D_{u0}=0$, but it is produced by V and therefore $u$ changes continuously in space. As the concentration of the nutrient $v_0$ is decreased for $D_0=0.05$, the localized state disappears and it decays to the uniform death state $u=w=0$ for $v_0\le 0.71$.  As $D_0$ is increased for $v_0=1$, the localized state disappears for $D_0\ge 0.121$ and leads to the death state $u=w=0$. Figure 1(b) displays the critical value $D_{0c}$ as a function of $v_0$.  The localized state is stable below the critical curve. The localized state cannot be maintained, if the diffusion constants are uniform $D_{v0}=D_{v1}$ and $D_{w0}=D_{w1}$. 
For $v_0\le 1.1$, $D_{0c}$ increases and the localized state is more stable for larger $v_0$. However, for $v_0>1.1$, $D_{0c}$ rapidly decreases. In this region, the localized state does not decay to the zero state $u=0$ but leads to an extended state for $D_0>D_{0c}$. An example of the extended state is shown in Fig.~1(c) at $v_0=1.3$ and $D_0=0.05$. Inside of the cell $r<r_0+d$, the equilibrium condition $u=v=w$ is almost satisfied.  We have found only the three types of solutions, i.e., the localized state, the uniform death state and the extended state, in our numerical simulation. Our localized nonequilibrium state is maintained by the outer boundary conditions. Blanchedeau et al. found spatial bistability of the thermodynamic state and the flow state in the chlorine-dioxide-iodide reaction in a thin film of gel fed from one side \cite{rf:13}. The phenomenon of the spatial bistability is also maintained by the supply of chemicals at the boundary. 

If we consider nonzero diffusivity at least for V and W,  the reactants and the waste materials are transported through the membrane and the nonequilibrium state is maintained. If $D_{v0}=0$ and $D_{u0}\ne 0,D_{w0}\ne 0$, the nutrient is not supplied through the membrane, and only the death state $u=w=0$ appears as shown in Fig.~2(a).  
If $D_{u0}=D_{w0}=0$ and $D_{v0}\ne 0$, the nutrient is supplied but the product and the waste material are not sent out from the cell. Then, an equilibrium state $u=v=w=v_0$ is attained for $0<r<r_0$ as shown in Fig.~2(b). If the diffusion constants $D_{u0},D_{v0}$ and $D_{w0}$ are all 0, the cell is completely isolated from the outside. In that case, the equilibrium state $u=v=w$ is attained inside of the cell as shown in Fig.~2(c). In this case, the final concentration of $u,v$ and $w$ depends on the initial concentration inside of the cell. In the simulation of Fig.~2(c), the initial concentrations were $u(r)=0.6, v(r)=0.5$ and $w(r)=0.4$ inside of the cell $r<r_0+d$. As a result of the conservation of the total concentration, the final concentration becomes the average value 0.5 of the initial concentrations.  
Thus, we see that the localized state shown in Fig.~1(a) is maintained only in the nonequilibrium condition.  

Even if $D_{v0}=D_{w0}=D_0<<1$ is not zero, the localized nonequilibrium state such as shown in Fig.~1(a) cannot be maintained, if the cell size $r_0+d$ is sufficiently large, because the central region of the cell is far away from the outside and an nearly equilibrium state is realized. Figure 3(a) displays the difference $v(0)-w(0)$ of the concentration as a function of $r_0$ for $D_0=0.05$ and $v_0=1$. The parameter $r_0$ is gradually increased from 1 to 4 for numerical simulations shown by rhombi. Then, $r_0$ is gradually decreased from 4 to 1, and the results of numerical simulations are denoted by crosses. If $r_0$ is larger than 3.2, the localized nonequilibrium state disappears abruptly and it leads to an extended state which is similar to the state shown in Fig.~3(a).  In the extended state, the equilibrium condition is nearly satisfied inside of the cell and therefore $v(0)-w(0)$ takes a very small value. The extended state is stable even for $r_0<3.2$, that is, the localized nonequilibrium state and the extended nearly equilibrium state are bistable as shown Fig.~3(a). The extended nearly equilibrium state at $v_0=1$, $r_0=2$ and $D_0=0.05$ is shown in Fig.~3(b), which is bistable with the localized state as shown in Fig.~1(a) at the same parameters. Our numerical result is that the localized nonequilibrium state exists only for small cell size $r_0+d$. The result suggests that the cell size cannot become too large in the primitive type metabolism, because the material transport occurs only through the simple diffusion. More elaborate active transport mechanisms such as a blood vessel network are necessary to maintain a large body.

 The localized state appears even if $D_{u0}$ is not zero. We consider hereafter the case of $D_u=D_v=D_w$ for the sake of simplicity. We assume $D_{u1}=D_{v1}=D_{v1}=1$, and $D_{u0}=D_{v0}=D_{w0}$ is expressed as $D_0$.  
  Figure 4(a) displays a localized state for $v_0=1$ and $D_0=0.02$. 
 Nonequilibrium active state is maintained inside of the cell, owing to the inward flow of the nutrient and the outward flow of the waste material.  
 Figure 4(b) displays the critical value $D_{0c}$ as a function of $v_0$.  The localized state is stable below the critical curve. 
The phase diagram is qualitatively the same as Fig.~1(b). 
The stable region of the localized state becomes smaller, because the self-catalytic product U can diffuse out.
When $D_u(r)=D_v(r)=D_w(r)=D(r)$ is satisfied, the sum $s=u+v+w$ obeys
\begin{equation}
\frac{\partial s}{\partial t}=\frac{1}{r^2}\frac{\partial}{\partial r}\left (Dr^2\frac{\partial u}{\partial r}\right ).
\end{equation}
The sum $s$ therefore approaches a uniform state $s(r)=v_0$, where $v_0$ is determined by the boundary condition. We have checked that the sum $s=u+v+w=v_0$ is actually satisfied in the stationary state shown in Fig.~4(a). 
The concentration $w$ is almost equal to $u$ as is seen in Fig.~4(a), but $w$ is slightly smaller than $u$. If $w=\alpha u$ ($\alpha\sim 1$) is assumed, $v=v_0-(1+\alpha)u$ is obtained. Substitution of these relations into eq.~(3) yields  
\begin{equation}
\frac{\partial u}{\partial t}=(v_0-1-\alpha)u^2-u^3-(1-\alpha)u+\frac{1}{r^2}\frac{\partial}{\partial r}\left (Dr^2\frac{\partial u}{\partial r}\right ).
\end{equation}
Figure 4(c) displays a localized solution of $u$ to eq.~(5) for $D_0=0.02$, $v_0=1$, $w=\alpha u$ and $v=1-u-\alpha u$. The parameter $\alpha=0.979$ is assumed, which is numerically obtained from $w(0)/u(0)$ in Fig.~4(a). 
The profiles of $u,v$ and $w$ in Fig.~4(c) are close to those shown in Fig.~4(a). 
It implies that the assumption $w=\alpha u$ is not exactly satisfied in the localized solution of eq.~(3), but the approximation is rather good.
\section{Localized solution in the Ginzburg-Landau equation with inhomogeneous diffusion constant}
To understand the mechanism of the stability of the localized  state, we study the simpler model (5) further. There are three uniform stationary solutions $u=0$ and $u=u_{0\pm}=(v_0-1-\alpha\pm\sqrt{(v_0-1-\alpha)^2-4(1-\alpha)})/2$ to eq.~(5), if $(v_0-1-\alpha)^2>4(1-\alpha)$. If $u/u_+$ is rewritten as a rescaled variable $u$, the corresponding reaction-diffusion equation becomes the Ginzburg-Landau type equation of the form: 
\begin{equation}
\frac{\partial u}{\partial t}=-u(u-a)(u-1)+\frac{1}{r^2}\frac{\partial}{\partial r}\left (Dr^2\frac{\partial u}{\partial r}\right )
\end{equation}
where $a=u_{0-}/u_{0+}$ and $D(r)=1$ for $r>r_0+d$ and $r<r_0$ and $D(r)=D_0$ for $r_0<r<r_0+d$.   There are two stable uniform solutions $u=0$ and $u=1$ for $0<a<1$ in this model. There is a potential function for this model:
\[E(\{u(r)\})=\int_0^{\infty}\left\{\frac{1}{4}u^4-\frac{1+a}{3}u^3+\frac{1}{2}au^2+\frac{1}{2}D(r)\left (\frac{\partial u}{\partial r}\right )^2\right \} 4\pi r^2 dr.\]
Equation (6) is a reaction diffusion equation with inhomogeneous diffusion constant. 
Several authors studied pinning phenomena of wave propagation in reaction diffusion systems with some inhomogeneities \cite{rf:14,rf:15} and discrete type reaction diffusion systems \cite{rf:16,rf:17}. A localized solution to eq.~(6) is related to these pinning phenomena. 

When the parameter $a=0.5$, the potential takes the same value for the two uniform solutions $u=0$ and $u=1$. Figure 5(a) is a localized state for $a=0.5,\,r_0=2,\,d=1$ and $D_0=0.015$. As $D_0$ is increased, the localized state disappears and it decays to the zero state $u=0$. We have investigated the critical value of $D_0$ above which the localized state disappears.   
Figure 5(b) displays the critical value $D_{0c}$ as a function of the thickness $d$ for $a=0.5$ and $r_0+d/2=5$. The localized state is most stable at $d\sim 1.5$, which is the optimum thickness of the membrane.  It is natural that localized activity in a cell with a thinner membrane is unstable, but it is not so trivial that the critical $D_{0c}$ decreases for large $d$. 
For $a<0.5$, and the potential effect makes the volume of $u=1$ increase, because the potential is lower for $u=1$.  If the radius $r_0$ is small, the curvature effect tends to make the inner volume of $u=1$ shrink. The localized solution is maintained by the pinning effect. The stability of the localized state is determined by the three effects.   
Figure 5(c) displays the value of $1-u$ at $r=r_0+d$  as a function of $r_0$ for $a=0.1,d=1$ and $D_0=0.05$. The localized solution disappears at $r_0=3$, when the radius $r_0$ is gradually increased from 1, and changed into the uniform state $u=1$. 
It is because the curvature effect becomes weak for larger $r_0$.  
The uniform state $u=1$ remains stable even for $r_0<3$, when $r_0$ is gradually decreased from $r_0=4$. 
This behavior is somewhat similar to Fig.~3(a).     
Figure 5(d) displays  the critical value of $D_0$ as a function of $r_0$ for $a=0.1$ and $d=1$. Under the critical curve, the localized solution is stable. For $r_0\le 1.9$, the localized solution changes into the uniform solution $u=0$, when $D_0$ is larger than $D_{0c}$. For $r_0\ge 1.95$, the localized solution changes into the uniform solution $u=1$, when $D$ is larger than $D_{0c}$.
This is because the curvature effect becomes too large for $r_0<1.9$ and the inner volume of $u=1$  shrinks for $D_0>D_{0c}$, even though the state of $u=1$ is preferable in the potential effect.

If $r_0$ is infinity, the problem is equivalent to a one-dimensional model.  Then, we study a simpler one-dimensional Ginzburg-Landau equation 
\begin{equation}
\frac{\partial u}{\partial t}=-u(u-a)(u-1)+\frac{\partial}{\partial x}\left (D\frac{\partial u}{\partial x}\right ).
\end{equation}
For the one-dimensional model, stationary solutions $u(x)$ can be explicitly solved, because $u(x)$ satisfies the equations:
\begin{eqnarray}
\frac{d^2u}{dx^2}&=&u(u-a)(u-1), \;\;{\rm for}\;\; x<r_0,\;x>r_0+d,\nonumber\\
D_0\frac{d^2u}{dx^2}&=&u(u-a)(u-1), \;\;{\rm for}\;\; r_0\le x\le r_0+d.
\end{eqnarray}
The solutions to eq.~(8) can be expressed using the elliptic functions.
The integral of eq.~(8) with respect to $x$ yields
\begin{eqnarray}
\frac{du}{dx}&=&-\sqrt{2\{E_1+u^4/4-(1+a)u^3/3+au^2/2\}}, \;\;{\rm for} \;\;x<r_0,\nonumber\\
\frac{du}{dx}&=&-\sqrt{(2/D_0)\{E_2+u^4/4-(1+a)u^3/3+au^2/2\}}, \;\;{\rm for} \;\;r_0\le x\le r_0+d,\nonumber\\
\frac{du}{dx}&=&-\sqrt{2\{E_3+u^4/4-(1+a)u^3/3+au^2/2\}}, \;\;{\rm for} \;\;x> r_0+d,
\end{eqnarray}
The constants $E_1$ and $E_3$ of the integral are determined as 
$E_1=-(2a-1)/12$ and $E_3=0$, because the solution approaches $u=1$ for $x\rightarrow -\infty$ and $u=0$ for $x\rightarrow \infty$. The constant $E_2$ must satisfy
\begin{equation}
E_2=\{u^4/4-(1+a)u^3/3+au^2/2\}(1/D_0-1)+E_1/D_0,
\end{equation}
 at $x=r_0$, because of the continuity of $D(x)(\partial u(x)/\partial x)$ at $x=r_0$.
The constant $E_2$ also needs to satisfy
\begin{equation} 
E_3=0=\{u^4/4-(1+a)u^3/4+au^2/2\}(D_0-1)+E_2D_0,
\end{equation} at $x=r_0+d$ because of the continuity of $D(x)(\partial u(x)/\partial x)$ at $x=r_0+d$.
\begin{figure}[t]
\begin{center}
\includegraphics[height=8cm]{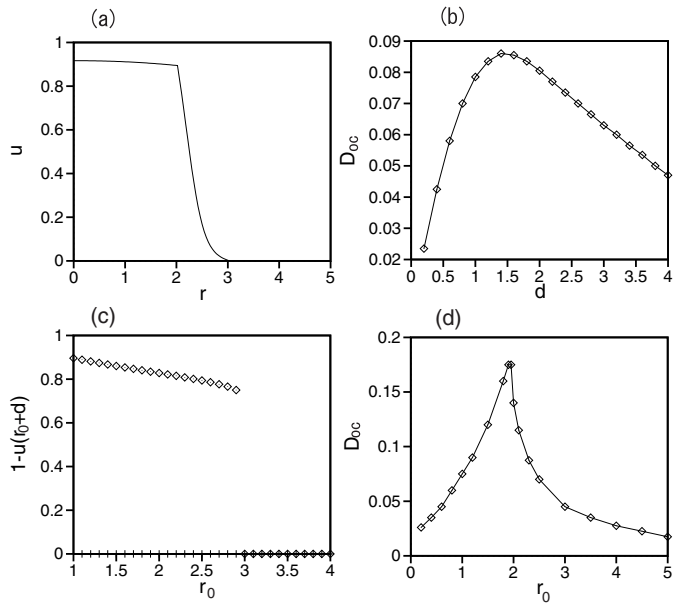}
\end{center}
\caption{(a) Stationary solution $u(r)$ to eq.~(6) for $a=0.5,\,r_0=2,\,d=1$ and $D_0=0.015$.  (b) Critical value $D_{0c}$ for the localized state as a function of $d$ at $a=0.5$ and $r_0+d/2=5$. (c) The value $1-u$ at the cell surface $r=r_0+d$  as a function of $r_0$ for $a=0.1,d=1$ and $D_0=0.05$. (d) Critical value $D_{0c}$ for the localized state as a function of $r_0$ at $a=0.1$ and $d=1$}
\label{f5}
\end{figure}
\begin{figure}[t]
\begin{center}
\includegraphics[height=5cm]{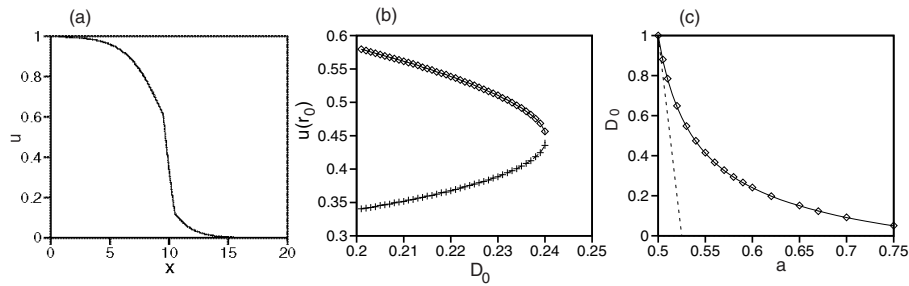}
\end{center}
\caption{(a) Stationary solution $u(r)$  for $a=0.6,\, d=1$ and $D_0=0.23$ by the direct numerical simulation of eq.~(7) (solid line) and the analysis by eq.~(8) (dashed line) (b) $u(r_0)$ as a function of $D_0$ for $a=0.6$ satisfying eq.~(10) and (11).  (c) Critical value $D_{0c}$ for the localized state as a function of $a$ at $d=1$. The dashed line is obtained from eq.~(12).}
\label{f6}
\end{figure}
We have numerically searched the value of $u(r_0)$ which satisfies the above conditions (10) and (11). If we can find the value of $u(r_0)$ at $x=r_0$, we can construct a localized solution by solving eq.~(9). Figure 6(a) compares the localized solution (dashed line) obtained by solving eq.~(9) using $u(r_0)=0.610335$ with the stationary solution (solid line) obtained by a direct numerical simulation of eq.~(7) for $a=0.6,\,d=1,x_0=9.5$ and $D_0=0.23$. The two curves overlap completely and we cannot distinguish the difference. There are discontinuity of the derivative $\partial u/\partial x$ at $x=r_0$ and $r_0+d$, because of the discontinuity of $D(x)$. 
We have numerically searched the value of $u(r_0)$ which satisfies the conditions (10) and (11) for various values of $D_0$. Figure 6(b) displays the obtained values of $u(r_0)$ as a function of $D_0$ for $a=0.6$. 
There are two solutions, and the upper branch represents a stable solution which appears in a direct numerical simulation. The lower one is an unstable solution. There is a saddle-node bifurcation at $D_0=D_{0c}\sim 0.235$. There is no localized solution for $D_0>D_{0c}$. The critical value $D_{0c}$ decreases with $a$ as shown in Fig.~6(c) for $d=1$. The rhombi denote the results by the direct numerical simulation and the solid line denotes the results of the numerical analysis for the saddle-node bifurcation. The two numerical results coincide well.  
The critical value $D_{0c}$ for $a<0.5$ is equal to $D_{0c}$ for $a^{\prime}=1-a>0.5$, owing to the symmetry of eq.~(7) with respect to the change of variable:$u\rightarrow 1-u$..  

As $a\rightarrow 0.5$, $D_{0c}$ approaches 1. In the limit of $a=0.5$ and $D_0=1$, the domain wall is expressed as $u=[1-{\rm tanh}\{\alpha(x-x_0)\}]/2$, where $\alpha=1/\sqrt{8}$ and $x_0$ denotes the position of the domain wall.  The potential is evaluated in the one-dimension system as 
\[E(x_0)=\int_{-\infty}^{\infty}\left\{\frac{1}{2}D(x)\left (\frac{\partial u}{\partial x}\right)^2+\frac{1}{4}u^4-\frac{1+a}{3}u^3+\frac{a}{2}u^2\right \}dx.\]
If $a-0.5<<1$, $1-D_0<<1$ and $x_0$ is close to $r_0+d/2$, the potential energy $E(x_0)$  is approximately given by 
\begin{equation}
E(x_0) \sim E(r_0+d/2)+\frac{2a-1}{12}(x_0-r_0-d/2)-\frac{\alpha(1-D_0)}{8}[{\rm tanh}\{\alpha(r_0+d-x_0)\}-{\rm tanh}\{\alpha(r_0-x_0)\}].
\end{equation}
The dashed line denotes the critical value  of $D_{0}$, above which a local minimum point disappears in the function $E(x_0)$. In other words, the domain wall can be trapped and pinned at the local minimum point of $E(x_0)$ by the inhomogeneity of $D(x)$, if $D_0$ is smaller than the critical value.  
The dashed line is tangential to the solid curve in the limit of $a=0.5$, but the two critical curves separate away rather rapidly, probably because the approximation of the domain wall by $u=[1-{\rm tanh}\{\alpha(x-x_0)\}]/2$ becomes worse as $a$ is increased.

\section{Summary and discussion}
We have performed numerical simulations of Schl\"ogl model in a vesicle with semi-permeable membrane as a primitive model of metabolism. High activity is maintained inside the cell, because of the low diffusivity at the membrane region. Absorption of nutrients and release of waste materials occur naturally as a result of the transport through the membrane, which is characteristic of the metabolism. The transport of material is essential for the metabolism, and  
it is characteristic of nonequilibrium state. The nonequilibrium state is maintained by the fixed boundary conditions $v=v_0$ and $w=0$ at $r=R$. 
Too large cell cannot maintain the nonequilibrium active state, because a nearly thermal equilibrium condition is satisfied deep inside the cell.

A theoretical analysis was performed for the Ginzburg-Landau equation with inhomogeneous diffusion constant to understand the mechanism of the stabilization of the localized state better. The Ginzburg-Landau equation was derived from the Schl\"ogl model using an assumption $w=\alpha u$, although the parameter $\alpha\sim w/u$ is not obtained theoretically. The mechanism of the localized state in the one dimensional model is interpreted as a kind of front pinning by the inhomogeneity. It might be an interesting result that there appears an optimal parameter value for the localized state with respect to the membrane width $d$ as shown in Fig.~5(b). The front pinning by inhomogeneity and discreteness were also studied by several authors, however, our main problem is the localized nonequilibrium state in the origin Schl\"ogl model including a thermal equilibrium state. The front pinning is a mechanism for the nonequilibrium localized state, but the problem should be further considered from a view point of nonequilibrium physics. For example, a localized state is possible even for a cell with large $r_0$ in the Ginzburg-Landau equation with inhomogeneous diffusion constant, if the parameter $a$ in eq.~(6) is assumed to be constant, but a localized nonequilibrium state is impossible in the original Schl\"ogl model for such a large cell.     

Our model might be too simple as a model for a living cell, because we do not consider complicated and realistic chemical reactions and various types of active transport of material through the membrane and inside of the cell. However, it might be an instructive model to consider metabolism and life conceptually from a view point of a dissipative structure far from equilibrium. It might be possible in the future to make an experiment of a primitive metabolism in an artificial cell. 


\begin{thebibliography}{99}
\bibitem{rf:1} G.~Nicolis and I.~Prigogine: {\it Self-Organization in Nonequilibrium Systems} (John Wiley and Sons, New York 1977) 
\bibitem{rf:2} Q.~Ouyang and H.~L.~Swinney: Nature {\bf 352} (1991) 1.
\bibitem{rf:3} A.~T.~Winfree: {\it The Geometry of Biological Time} (Springer-Verlag, Berlin 1980).
\bibitem{rf:4} P.~Gray and S.~K.~Scott: J. Phys. Chem. {\bf 89} (1985) 22.
\bibitem{rf:5} J.~Pearson: Science {\bf 261} (1993) 189. 
\bibitem{rf:6} K-J.~Lee, W.~D.~McCormick, J.~E.~Pearson and H.~L.~Swinney: Nature {\bf 369} (1994) 215.
\bibitem{rf:7} D.~Winston, M.~Arora, J.~Maselko, V.~Gaspar, and K.~Showalter: Nature {\bf 351} (1991) 132.
\bibitem{rf:8} V.~K.~Vanag and I.~R.~Epstein: Phys. Rev. Letts. {\bf 87} (2001) 228301.
\bibitem{rf:9} A.~I.~Oparin: {\it The Origin of Life},   Dover, New York (1952).
\bibitem{rf:10} F.~J.~Dyson: {\it Origin of Life}, Cambridge University Press, Cambridge (1985).
\bibitem{rf:11} V.~Noireaux and A.~Libchaber: Proc. Nat. Acad. Sci {\bf 101} (2004) 17669.
\bibitem{rf:12} F.~Schl\"ogl: Z.Physik {\bf 248} (1971) 446.
\bibitem{rf:13} P.~Blanchedeau, J.~Boissonade, and P.~De Kepper: Physica D {\bf 147} (2000) 283.
\bibitem{rf:14} A.~Kulka, M.~Bode and H.~G.~Purwins: Phys. Lett. A {\bf 203} (1995) 33.
\bibitem{rf:15} Y.~Nishiura, M.~Mimura, H.~Ikeda and H.~Fujii: SIAM J. Math. Anal. {\bf 21} (1990) 85. 
\bibitem{rf:16} I.~Mittkov, K.~Kladko and J.~E.~Pearson: Phys. Rev. Lett. {\bf 81} (1998) 5453.
\bibitem{rf:17} A.~Carpio and L.~L.~Bonilla: Phys. Rev. Lett. {\bf 86} (2001) 6034.
\end{thebibliography}
\end{document}